\documentclass[12pt]{article}
\usepackage{amsmath}
\usepackage{amssymb}
\usepackage{graphicx}

\input{tcilatex}
\begin{document}

\begin{center}
\textbf{Higher Dimensional Cosmology: Relations among the radii of two
homogeneous spaces.}\bigskip

\bigskip

\bigskip

\smallskip\ 

E. A. Le\'{o}n$^{\ast \ddag }$ \footnote{%
ealeon@posgrado.cifus.uson.mx}, J. A. Nieto$^{\star \ast \dag }$ \footnote{%
nieto@uas.uasnet.mx}, R. N\'{u}\~{n}ez-L\'{o}pez$^{\ast }$ \footnote{%
ramona@cifus.uson.mx} and A. Lipovka$^{\ast }$ \footnote{%
aal@cajeme.cifus.uson.mx}

\smallskip

$^{\ast }$\textit{Departamento de Investigaci\'{o}n en F\'{\i}sica de la
Universidad de Sonora, 83000, Hermosillo Sonora , M\'{e}xico}\ 

$^{\star }$\textit{Facultad de Ciencias F\'{\i}sico-Matem\'{a}ticas de la
Universidad Aut\'{o}noma} \textit{de Sinaloa, 80010, Culiac\'{a}n Sinaloa, M%
\'{e}xico.}

$\dag $\textit{Mathematical, Computational \& Modeling Science Center,
Arizona State University, PO Box 871904, Tempe, AZ 85287, USA}

$\ddag $\textit{Universidad Aut\'{o}noma de Sinaloa, Facultad de Ingenier%
\'{\i}a Mochis, 81220, Los Mochis Sinaloa, M\'{e}xico}

\ 

\bigskip\ 

\textbf{Abstract}
\end{center}

\noindent We study a cosmological model in 1+D+d dimensions where D
dimensions are associated with the usual Friedman-Robertson-Walker type
metric with radio a(t) and d dimensions corresponds to an additional
homogeneous space with radio b(t). We make a general analysis of the field
equations and then we obtain solutions involving the two cosmological radii,
a(t) and b(t). The particular case D=3, d=1 is studied in some detail.\ \ 

\bigskip\ 

\bigskip\ 

\bigskip\ 


Pacs numbers: 04.50.-h, 04.50.Cd, 98.80.-k

December, 2010

\newpage \noindent \textbf{I. Introduction.}\bigskip

\noindent Extra-dimensional models have been subject of interest since its
appearance in physics, mainly as a basis for unification models [1]-[4]. In
fact, although there has been renewed interest in such models through the
development of string theory, the very first works on the subject (Nordstr%
\"{o}m, Kaluza and Klein)[1] contained already the essential techniques for
building theories in higher dimensions, particularly the idea of
compactification.

Different approaches to the problem eventually arose (two nice reviews are
[1] and [2]), as well as considerations regarding the possibility of
noncompact extra dimensions [5][6] or any type of dynamical compactification
[7]-[11]. All possibilities studied had ultimately as common point the
cosmological implications of the theory. This is because the cosmological
scales may amplify any particular effect of a model with extra dimensions.

However, the extra-dimensional cosmological model appropiate to any
particular approach relies on some assumptions by itself. This in turn
permit us to observe that the study of cosmological models with extra
dimensions deserves atention by itself. In this context, we review a
cosmological model with extra dimensions, as it was presented in Ref. [12],
such that $D$ dimensions have an evolving radius $a$ and $d$ dimensions have
another evolving radius $b$. Specifically, we consider a higher dimensional
cosmological model with metric [12][13]%
\begin{equation}
ds^{2}=-dt^{2}+a^{2}(t)\tilde{g}_{ij}(x)dx^{i}dx^{j}+b^{2}(t)\hat{g}%
_{ab}(y)dy^{a}dy^{b}.  \tag{1.1}
\end{equation}%
Here, indices ($i$, $j$, ...) run from $1$ to $D$, while ($a$, $b$, ...) run
from $D+1$ to $D+d$. Also, $\tilde{g}_{ij}(x)$ and $\hat{g}_{ab}(y)$ are the
metrics for two homogeneous spaces that depend only on the co-moving
coordinates $(x^{1},...,x^{D})$ and $(y^{D+1},...,y^{D+d})$, respectively.
From here on, these spaces will be called $D$-space and $d$-space, with an
obvious connotation.

With the prescriptions given above, Einstein equations in vacuum (cf.
Appendix) are 
\begin{equation}
D(D-1)\frac{\dot{a}^{2}}{a^{2}}+D(D-1)\frac{k_{1}}{a^{2}}+d(d-1)\frac{\dot{b}%
^{2}}{b^{2}}+d(d-1)\frac{k_{2}}{b^{2}}+2dD\frac{\dot{a}\dot{b}}{ab}=0; 
\tag{1.2}
\end{equation}%
\begin{equation}
\begin{array}{c}
(D-1)\left\{ \frac{\ddot{a}}{a}+\frac{(D-2)}{2}\left[ \frac{\dot{a}^{2}}{%
a^{2}}+\frac{k_{1}}{a^{2}}\right] \right\} +d\left\{ \frac{\ddot{b}}{b}+%
\frac{(d-1)}{2}\left[ \frac{\dot{b}^{2}}{b^{2}}+\frac{k_{2}}{b^{2}}\right]
\right\} \\ 
\\ 
+d(D-1)\frac{\dot{a}\dot{b}}{ab}=0;%
\end{array}
\tag{1.3}
\end{equation}

\begin{equation}
\begin{array}{c}
(d-1)\left\{ \frac{\ddot{b}}{b}+\frac{(d-2)}{2}\left[ \frac{\dot{b}^{2}}{%
b^{2}}+\frac{k_{2}}{b^{2}}\right] \right\} +D\left\{ \frac{\ddot{a}}{a}+%
\frac{(D-1)}{2}\left[ \frac{\dot{a}^{2}}{a^{2}}+\frac{k_{1}}{a^{2}}\right]
\right\} \\ 
\\ 
+D(d-1)\frac{\dot{a}\dot{b}}{ab}=0.%
\end{array}
\tag{1.4}
\end{equation}

The structure of this article is as follows: In this section (I) we have
presented the model, and shown the corresponding Einstein equations. Next
two sections are focused in the particular case of $3+1$ spatial dimensions.
In section II the vacuum case is presented briefly, while Section III deals
with the inclusion of matter. We solve there $a$ as function of time, as
well as $b$ as function of $a$. Section III ends with the general solutions
when $\Lambda \neq 0$. Then, we present in section IV an interesting result
where radii $a$ and $b$ are related in a form that resembles a conserved
"angular momentum". Finally, we present a summary of the work and make some
final comments in section V.

In resume, our main results of this study are: first, we obtain the
behaviour of $a(t)$ and $b(a)$ for the two cases in which the cosmological
constant is zero or nonzero, and secondly we show a general relation between
the two radii in the presence of matter, reminiscent of the conservation of
classical angular momentum. We start the analysis with a special case, where 
$D=3$ and $d=1$ [14][15].

\bigskip

\noindent \textbf{II. D=3, d=1 dimensional model.}\bigskip

\noindent By setting $D=3$ and $d=1$, equations (1.2)-(1.4) are, respectively%
\begin{equation}
\frac{\dot{a}^{2}}{a^{2}}+\frac{k_{1}}{a^{2}}+\frac{\dot{a}\dot{b}}{ab}=0, 
\tag{2.1}
\end{equation}%
\begin{equation}
2\frac{\ddot{a}}{a}+\frac{\dot{a}^{2}}{a^{2}}+\frac{k_{1}}{a^{2}}+\frac{%
\ddot{b}}{b}+2\frac{\dot{a}\dot{b}}{ab}=0  \tag{2.2}
\end{equation}%
and%
\begin{equation}
\frac{\ddot{a}}{a}+\frac{\dot{a}^{2}}{a^{2}}+\frac{k_{1}}{a^{2}}=0. 
\tag{2.3}
\end{equation}

Equation (2.3) can be integrated respect to $a$ by mean of a change of
variable [14], resulting 
\begin{equation}
\dot{a}^{2}+k_{1}=\frac{\Gamma }{a^{2}},  \tag{2.4}
\end{equation}%
where $\Gamma $ is a integration constant. Equation (2.4) serves two
purposes. The first is to give the behaviour of $a$ in time. Second, it
permit us to integrate (2.1) in order to obtain radius $b$ as a function of $%
a$, even with the inclusion of matter. We review both results more
extensively in next section. By the moment, we verify that in this case
-where $\rho =0$-, we obtain $b$ as function of $a$ directly from (2.1) and
(2.3), since both together imply $\frac{\ddot{a}}{a}=\frac{\dot{a}\dot{b}}{ab%
}$, and this in turn yields 
\begin{equation}
b=\beta \dot{a},  \tag{2.5}
\end{equation}%
where $\beta $ is a positive constant for a expanding Universe in $3$-space.
We mention that, by virtue of (2.3)-(2.5) we have that%
\begin{equation}
\dot{b}=-\frac{\beta \Gamma }{a^{3}},  \tag{2.6}
\end{equation}%
i.e., with $\beta $ and $\Gamma >0$ we have a decreasing radius $b$ in time
for this vacuum case. As we will see, all this results are limit cases of a
more general result, when matter is included.\bigskip 

\noindent \textbf{III. D=3, d=1 model with matter.}\bigskip

\noindent In this section, we review a Kaluza-Klein cosmological model where 
$D=3$ and $d=1$, with a "five-dimensional dust", as seen in [14]. In this
case, (A.6) of the Appendix, where $T_{00}$ corresponds to $\rho $, the
five-dimensional density, can be written as 
\begin{equation}
\frac{\dot{a}^{2}}{a^{2}}+\frac{k_{1}}{a^{2}}+\frac{\dot{a}\dot{b}}{ab}=%
\frac{8\pi \mu }{3a^{3}b}.  \tag{3.1}
\end{equation}%
Here, Bianchi identities $G_{\quad ;\mu }^{\mu \nu }=0$ gave $\rho =\frac{%
\mu }{a^{3}b}$, were $\mu $ is a constant.

First, from equation (2.4) we obtain the behaviour of radius $a$ for the
cases of zero ($k_{1}=0$), positive ($k_{1}=1$) or negative curvature ($%
k_{1}=-1$) in 3-space:

a) $k_{1}=0$. In this case $\Gamma $ is necessarily positive, and the
integration yields 
\begin{equation}
a=\sqrt{2\sqrt{\Gamma }t}.  \tag{3.2}
\end{equation}%
This gives the evolution of the Hubble parameter as $H=\frac{1}{2t}=\frac{%
\sqrt{\Gamma }}{a^{2}}$. The result can be constrasted with that of the FRW
model in a matter dominated Universe, where $a\propto t^{\frac{2}{3}}$, $%
H\propto a^{-\frac{3}{2}}$, and $H\propto \frac{1}{t}$ [16].

b) $k_{1}=1$. Here $\Gamma $ is strictly positive too, and $a(t)$ is 
\begin{equation}
a=\left[ \Gamma -(\sqrt{\Gamma }-t)^{2}\right] ^{\frac{1}{2}}.  \tag{3.3}
\end{equation}

c) $k_{1}=-1$. Here $\dot{a}=\frac{\sqrt{\Gamma +a^{2}}}{a}$ and $\Gamma $
can be positive or negative. For $\Gamma >0$, we have%
\begin{equation}
a=\left[ (t+\sqrt{\Gamma })^{2}-\Gamma \right] ^{\frac{1}{2}}.  \tag{3.4}
\end{equation}%
Thus, the qualitative picture is a deccelerating (expanding-) Universe.

If $\Gamma <0$, the behaviour of $a$ in time is given by%
\begin{equation}
a=\left[ t^{2}-\Gamma \right] ^{\frac{1}{2}},  \tag{3.5}
\end{equation}%
where we set $t=0$ as the time when $a$ reaches a minimum $a_{\min }=\sqrt{%
-\Gamma }$ (units where $c=1$). We stress the fact that this solution
doesn't need the introduction of a cosmological constant for an accelerating
radius $a$.

Fig. 1 shows the behaviour of a(t) for the different cases just analysed.

\begin{figure}[htbp]
\includegraphics[scale=0.4]{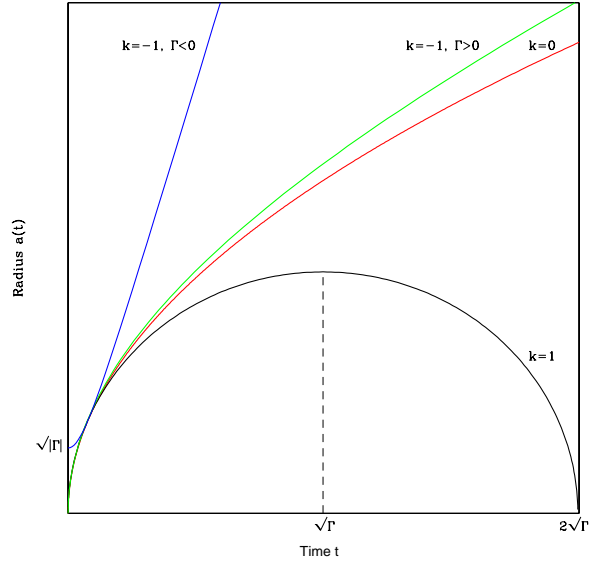} 
\label{Fig. 1}\centering
\caption{{\protect\footnotesize Evolution of radius }}$a(t)${\footnotesize \ in
time for the three cases of }$k${\footnotesize \ (-1, 0 and +1). The fourth
case, with }$\Gamma <0${\footnotesize \ is shown in blue.}
\end{figure}

\bigskip 

It is interesting to note that the mere introduction of the extra dimension
-even in vacuum- gives the Friedmann type equation (2.4), and then we can
consider an effective density $\rho _{eff}=\frac{3\Gamma }{8\pi a^{4}}$. In
the classical FRW model this density corresponds to ultra-relativistic
matter, related with pressure by $p=\rho /3$. Then in this model $\Gamma <0$
corresponds to a solution with an effective negative pressure. We note that
according to (2.1), we may also identify the term $\frac{\dot{a}\dot{b}}{ab}$
as the effective density (more precisely $\frac{\dot{a}\dot{b}}{ab}=-\frac{%
8\pi }{3}\rho _{eff}$).

By using (2.4) and (3.1), we obtain%
\begin{equation}
\frac{\Gamma }{a^{4}}+\frac{\Gamma -k_{1}a^{2}}{a^{3}b}\frac{db}{da}=\frac{%
8\pi \mu }{3a^{3}b}.  \tag{3.6}
\end{equation}

This can be integrated to obtain $b=b(a)$. We have three different cases:%
\begin{equation}
k=0\rightarrow b=\frac{4\pi \mu a}{3\Gamma }+\frac{\alpha }{a},  \tag{3.7}
\end{equation}%
with $\alpha =const$. Observe also that making the identification $\alpha
=\beta \sqrt{\Gamma }$ in (3.7) we obtain (2.5) as limit case when $\rho
\rightarrow 0$ (contrast for instance with Ref. [3]).%
\begin{equation}
k=1\rightarrow b=\frac{8\pi \mu }{3a}+\beta \dot{a},  \tag{3.8}
\end{equation}%
where $\dot{a}=\frac{\sqrt{\Gamma -a^{2}}}{a}$ and $\beta $ is a constant
consistent with (2.5). The third case is%
\begin{equation}
k=-1\rightarrow b=-\frac{8\pi \mu }{3a}+\beta \dot{a}.  \tag{3.9}
\end{equation}%
Here, $\dot{a}=\frac{\sqrt{a^{2}+\Gamma }}{a}$ and $\Gamma $ can be positive
or negative as we saw before.

We can see that the set of solutions for $b(a)$ given in (3.7)-(3.9) yield
for $\mu \rightarrow 0$ the limit value $b\rightarrow \beta \dot{a}$, in
agreement with (2.5). We show the schematic evolution of $b$ in time in Fig.
2 for the four cases considered.

\begin{figure}[htbp]
\includegraphics[scale=0.4,keepaspectratio=true]{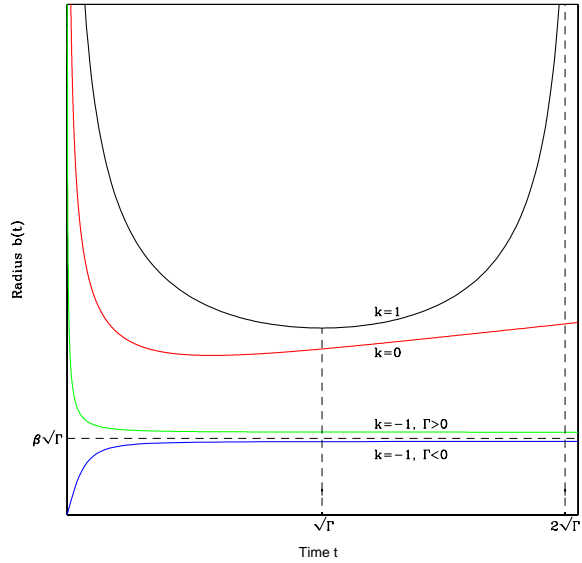} 
\label{Fig. 2}\centering
\caption{{\protect\footnotesize Schematic evolution of \ radius }}$b(t)$%
{\footnotesize .}
\end{figure}

\bigskip 

Now consider a nonvanishing cosmological constant; in this case, in the LHS
of equations (2.3) and (3.1) results an extra term $-\frac{\Lambda }{3}$.
The analogue of (2.4) turns out to be 
\begin{equation}
\dot{a}^{2}+k_{1}=\frac{\Lambda }{6}a^{2}+\frac{\Gamma }{a^{2}}.  \tag{3.10}
\end{equation}

Solving for $a(t)$, we obtain 
\begin{equation}
a^{2}=\frac{3k}{\Lambda }+\frac{M}{2}e^{\sqrt{\frac{2\Lambda }{3}}\Delta t}+%
\frac{1}{M}\left( \frac{9k^{2}}{2\Lambda ^{2}}-\frac{3\Gamma }{\Lambda }%
\right) e^{-\sqrt{\frac{2\Lambda }{3}}\Delta t},  \tag{3.11}
\end{equation}%
where we have defined $M=\left( a_{0}^{2}-\frac{3k}{\Lambda }\right) +\sqrt{%
\left( a_{0}^{2}-\frac{3k}{\Lambda }\right) ^{2}+\frac{6\Gamma }{\Lambda }-%
\frac{9k^{2}}{\Lambda ^{2}}}$; also $\Delta t=t-t_{0}$ and $a_{0}=a(t_{0})$.
Fig. 3 shows the \ schematic behaviour for the evolution of $a(t)$.

\begin{figure}[htbp]
\includegraphics[scale=0.4]{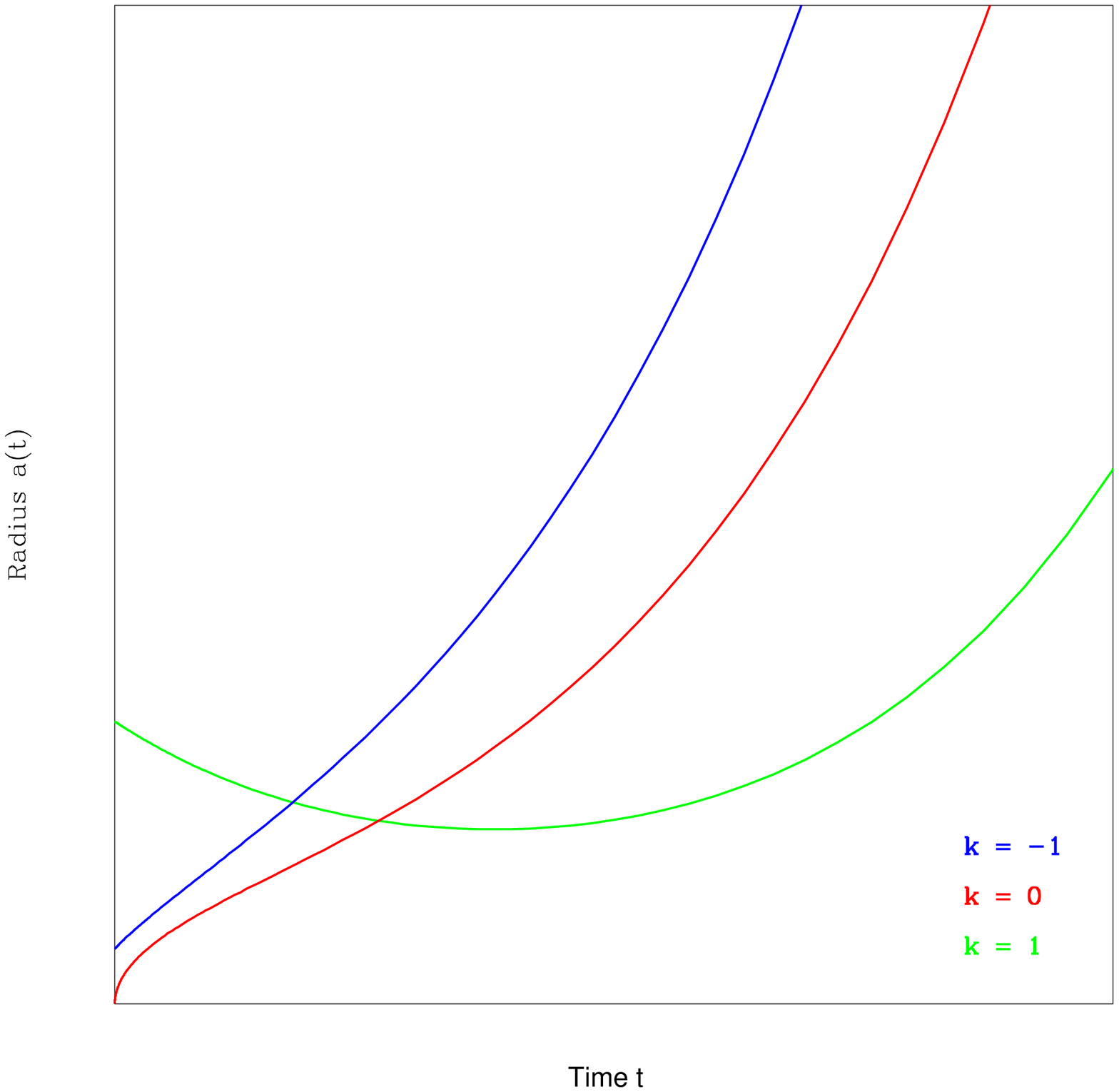} 
\label{Fig. 3}\centering
\caption{{\protect\footnotesize Evolution of radius }}
$a(t)${\footnotesize \ in time when }$\Lambda \neq 0${\footnotesize .}
\end{figure}

\bigskip

By using (3.10) and the first Einstein equation with cosmological constant,
we have%
\begin{equation}
\frac{\Gamma }{a^{4}}+\frac{\dot{a}\dot{b}}{ab}-\frac{\Lambda }{6}=\frac{%
8\pi \mu }{3a^{3}b}.  \tag{3.12}
\end{equation}%
Taking $b$ as function of $a$, the result is%
\begin{equation}
b=\frac{8\pi \mu \left( \Lambda a^{2}-3k\right) }{3\left( 2\Gamma \Lambda
-3k^{2}\right) a}+\beta \frac{\sqrt{\Gamma +\frac{\Lambda }{6}a^{4}-ka^{2}}}{%
a}.  \tag{3.13}
\end{equation}%
We have used the same integration constant $\beta $ stressing the fact that $%
\Lambda =0$ reduces to the previous cases.

\bigskip\ 

\noindent \textbf{IV. (1+D+d)-dimensional model.}\bigskip

\noindent Now we turn to the general case $D+d+1$. In this case it cannot be
defined exact solutions for the model, except for some assumptions (e.g. see
Ref. [13]). Here, we remark a general relation between $a(t)$ and $b(t)$. We
define three functions of time $F$, $G$ and $H$, given by

\begin{equation}
F\equiv \frac{\ddot{a}}{a}+(D-1)\left( \frac{\dot{a}^{2}+k_{1}}{a^{2}}%
\right) +d\frac{\dot{a}\dot{b}}{ab},  \tag{4.1}
\end{equation}%
\begin{equation}
G\equiv \frac{\ddot{b}}{b}+(d-1)\left( \frac{\dot{b}^{2}+k_{2}}{b^{2}}%
\right) +D\frac{\dot{a}\dot{b}}{ab}  \tag{4.2}
\end{equation}%
and%
\begin{equation}
H\equiv D\left[ \frac{\ddot{a}}{a}+\frac{(D-1)}{2}\left( \frac{\dot{a}%
^{2}+k_{1}}{a^{2}}\right) \right] +d\left[ \frac{\ddot{b}}{b}+\frac{(d-1)}{2}%
\left( \frac{\dot{b}^{2}+k_{2}}{b^{2}}\right) \right] +dD\frac{\dot{a}\dot{b}%
}{ab}.  \tag{4.3}
\end{equation}%
In terms of them, (1.2) can be written as%
\begin{equation}
D\left[ F-\frac{\ddot{a}}{a}\right] +d\left[ G-\frac{\ddot{b}}{b}\right] =0.
\tag{4.4}
\end{equation}%
Also, (1.3) and (1.4) can be written respectively as 
\begin{equation}
H-F=0  \tag{4.5}
\end{equation}%
and 
\begin{equation}
H-G=0.  \tag{4.6}
\end{equation}%
Both equations imply that 
\begin{equation}
F=G.  \tag{4.7}
\end{equation}

Now we consider a perfect fluid in higher dimensions, such that the
energy-momentum tensor takes the form 
\begin{equation}
T_{\mu \nu }=(p+\rho )U_{\mu }U_{\nu }+pg_{\mu \nu };  \tag{4.8}
\end{equation}%
here $p$ and $\rho $ are the pressure and the density of the fluid
respectively, and $U_{\mu }$ is the $n-$velocity vector, which has component 
$U_{0}=-1$ in this comoving coordinate frame defined by (1.1).

It is easy to show that the first Einstein equation, namely (A.6) of the
Appendix, is%
\begin{equation}
D\left[ F-\frac{\ddot{a}}{a}\right] +d\left[ G-\frac{\ddot{b}}{b}\right]
=8\pi T_{00}.  \tag{4.9}
\end{equation}%
Also, (A.7) and (A.8) corresponds to%
\begin{equation}
H-F=-8\pi p  \tag{4.10}
\end{equation}%
and%
\begin{equation}
H-G=-8\pi p,  \tag{4.11}
\end{equation}%
respectively, implying that (4.7) still holds. If we assume (4.9) to be
solvable by%
\begin{equation}
F-\frac{\ddot{a}}{a}=\frac{8\pi }{D+d}\rho  \tag{4.12}
\end{equation}%
and%
\begin{equation}
G-\frac{\ddot{b}}{b}=\frac{8\pi }{D+d}\rho ,  \tag{4.13}
\end{equation}%
by combining (4.7), (4.12) and (4.13), then results%
\begin{equation}
a\dot{b}-b\dot{a}=const.,  \tag{4.14}
\end{equation}%
which is analogue to the definition of classical angular momentum.\bigskip
\bigskip

\noindent \textbf{V. Final comments.}\bigskip

\noindent In this work we studied a general model with $D$ dimensions
related by the same evolving radius $a$, and $d$ dimensions with evolving
radius $b$. From the introduction of the vacuum case with 3+1 spatial
dimensions in section II, we have shown in section III the evolution in time
of $a$ and $b$ for the case of a five-dimensional dust. In this part we have
followed closely [9] and we have included the case with non-vanishing
cosmological constant. We can mention that in the process we have obtained
the correct expression (3.7) for $b=b(a)$ in the flat case $k=0$ -instead of
the second expression in equation (2.6) of Reference [9]- that gives as
limiting case (2.5). Also, we remark again the special case $k<0$, $\Gamma
<0 $ where the solution implies an accelerated rate of expansion for $a(t)$,
without the introduction of a cosmological constant, for instance. Further,
we have obtained the general solutions when $\Lambda $ is taken into
account, both for $a(t)$ and $b(a)$. In section IV we have obtained an
interesting relation in he form of eq. (4.14) for radii $a$ and $b$, valid
for a special distribution of mater in the two homogeneous spaces. Our
analysis can be complementary in the study of dynamical compactification
[10]-[12] and in general for extra dimensional models [15][17].

\bigskip

\begin{center}
\textbf{Acknowledgments}
\end{center}

E. A. Le\'{o}n gratefully acknowledge a PhD fellowship by CONACyT. J.A. Nieto would like to thank both the Departamento de Investigaci\'{o}n en F\'{\i}sica de la Universidad de Sonora and the Mathematical, Computational and Modeling Science Center at the Arizona State University for the hospitality, where part of this work was developed.

\bigskip

\noindent \textbf{APPENDIX }\bigskip

\noindent \textbf{Einstein equations}\bigskip\ 

\noindent Here, we follow closely Ref. [5]. According to the notation in the
text, from (1.1) we obtain the nonvanishing Christoffel symbols $\Gamma
_{\alpha \beta }^{\mu }=\frac{1}{2}g^{\mu \lambda }(g_{\alpha \lambda ,\beta
}+g_{\beta \lambda ,\alpha }-g_{\alpha \beta ,\lambda })$,%
\begin{equation}
\begin{array}{ccc}
\Gamma _{ij}^{0}=a\dot{a}\tilde{g}_{ij}, & \Gamma _{0j}^{i}=\frac{\dot{a}}{a}%
\delta _{j}^{i}, & \Gamma _{jk}^{i}=\tilde{\Gamma}_{jk}^{i}, \\ 
&  &  \\ 
\Gamma _{ab}^{0}=b\dot{b}\hat{g}_{ab} & \Gamma _{0b}^{a}=\frac{\dot{b}}{b}%
\delta _{b}^{a}, & \Gamma _{bc}^{a}=\hat{\Gamma}_{bc}^{a};%
\end{array}
\tag{A.1}
\end{equation}%
where $\tilde{\Gamma}_{jk}^{i}$ and $\hat{\Gamma}_{bc}^{a}$ refers to
Christoffel symbols in the adequate dimensional reduction. We use the same
notation as indicative of dimensional reduction; for instance, just as $%
\tilde{\Gamma}_{jk}^{i}=\frac{1}{2}g^{il}(g_{jl,k}+g_{lk,j}-g_{jk,l})$, we
have $\hat{R}_{\ bcd}^{a}=\partial _{c}\hat{\Gamma}_{db}^{a}-\partial _{d}%
\hat{\Gamma}_{cb}^{a}+\hat{\Gamma}_{cf}^{a}\hat{\Gamma}_{db}^{f}-\hat{\Gamma}%
_{df}^{a}\hat{\Gamma}_{cb}^{f}$ and $\tilde{R}=\tilde{g}^{ij}\tilde{R}_{ij}$%
. After calculating the needed components of the Riemann tensor $R_{\ \nu
\alpha \beta }^{\mu }=\partial _{\alpha }\Gamma _{\beta \nu }^{\mu
}-\partial _{\beta }\Gamma _{\alpha \nu }^{\mu }+\Gamma _{\alpha \lambda
}^{\mu }\Gamma _{\beta \nu }^{\lambda }-\Gamma _{\beta \lambda }^{\mu
}\Gamma _{\alpha \nu }^{\lambda }$ and contracting it as $R_{\mu \nu }=$ $%
R_{\ \mu \alpha \nu }^{\alpha }$, we obtain three classes for the non-zero
Ricci tensor components: for the time component,%
\begin{equation}
R_{00}=-D\frac{\ddot{a}}{a}-d\frac{\ddot{b}}{b};  \tag{A.2}
\end{equation}%
for the D dimensions $(x^{i},x^{j},...)$, 
\begin{equation}
R_{ij}=\tilde{R}_{ij}+\tilde{g}_{ij}\left[ a\ddot{a}+(D-1)\dot{a}^{2}+d\frac{%
a}{b}\dot{a}\dot{b}\right] ;  \tag{A.3}
\end{equation}%
and for the d dimensions $(x^{a},x^{b},...)$ 
\begin{equation}
R_{ab}=\hat{R}_{ab}+\hat{g}_{ab}\left[ b\ddot{b}+(d-1)\dot{b}^{2}+D\frac{b}{a%
}\dot{a}\dot{b}\right] .  \tag{A.4}
\end{equation}%
With another contraction, the curvature scalar is 
\begin{equation}
R=2D\frac{\ddot{a}}{a}+2d\frac{\ddot{b}}{b}+\frac{1}{a^{2}}[\tilde{R}+D(D-1)%
\dot{a}^{2}+dD\frac{a}{b}\dot{a}\dot{b}]+\frac{1}{b^{2}}[\hat{R}+d(d-1)\dot{b%
}^{2}+dD\frac{b}{a}\dot{a}\dot{b}].  \tag{A.5}
\end{equation}

Of course, all the results calculated above are symmetric respect to the
D-dimensional and d-dimensional subspaces defined by the metric (1.1). More
precisely, we obtain the same tensors if we perform the interchage $%
a\leftrightarrow b$ and $D\leftrightarrow d$, as well as the changes they
induce, v.g. $\tilde{R}\leftrightarrow $ $\hat{R}$. Furthermore, due to the
homogeneity in the mentioned subspaces, we can write $\tilde{R}%
_{ij}=k_{1}(D-1)\tilde{g}_{ij}$ and $\hat{R}_{ab}=k_{2}(d-1)\hat{g}_{ab}$.,
were $k_{1}$ and $k_{2}$ can take values in $\{-1,0,1\}$.

With all this at hand, -working in units where $c=G=1$- we can express the
following sets of Einstein equations, $G_{\mu \nu }=R_{\mu \nu }-\frac{1}{2}%
g_{\mu \nu }R=8\pi T_{\mu \nu }$:%
\begin{equation}
\begin{array}{c}
G_{00}=8\pi T_{00}\rightarrow \\ 
\\ 
\frac{D(D-1)}{2}\left( \frac{\dot{a}^{2}+k_{1}}{a^{2}}\right) +\frac{d(d-1)}{%
2}\left( \frac{\dot{b}^{2}+k_{2}}{b^{2}}\right) +dD\frac{\dot{a}\dot{b}}{ab}%
=8\pi T_{00};%
\end{array}
\tag{A.6}
\end{equation}%
\begin{equation}
\begin{array}{c}
G_{ij}=8\pi T_{ij}\rightarrow \{(D-1)\left[ \frac{\ddot{a}}{a}+\frac{(D-2)}{2%
}\left( \frac{\dot{a}^{2}+k_{1}}{a^{2}}\right) \right] \\ 
\\ 
+d\left[ \frac{\ddot{b}}{b}+\frac{(d-1)}{2}\left( \frac{\dot{b}^{2}+k_{2}}{%
b^{2}}\right) \right] +d(D-1)\frac{\dot{a}\dot{b}}{ab}\}a^{2}\tilde{g}%
_{ij}=-8\pi T_{ij};%
\end{array}
\tag{A.7}
\end{equation}%
\begin{equation}
\begin{array}{c}
G_{ab}=8\pi T_{ab}\rightarrow \{(d-1)\left[ \frac{\ddot{b}}{b}+\frac{(d-2)}{2%
}\left( \frac{\dot{b}^{2}+k_{2}}{b^{2}}\right) \right] \\ 
\\ 
+D\left[ \frac{\ddot{a}}{a}+\frac{(D-1)}{2}\left( \frac{\dot{a}^{2}+k_{1}}{%
a^{2}}\right) \right] +D(d-1)\frac{\dot{a}\dot{b}}{ab}\}b^{2}\hat{g}%
_{ab}=-8\pi T_{ab}.%
\end{array}
\tag{A.8}
\end{equation}

\bigskip\

\end{document}